\documentclass[journal]{IEEEtran}
\usepackage{amsmath,amsfonts}
\usepackage{algorithmic}
\usepackage{algorithm}
\usepackage{array}
\usepackage[caption=false,font=scriptsize,labelfont=sf,textfont=sf]{subfig}
\usepackage{textcomp}
\usepackage{stfloats}
\usepackage{url}
\usepackage{verbatim}
\usepackage{graphicx}
\graphicspath{{figures}{figs}}
\usepackage[capitalize]{cleveref}
\usepackage{cite}
\usepackage{color}
\usepackage[normalem]{ulem}
\newcommand{\added}[1]{\textcolor{blue}{\uwave{#1}}}
\newcommand{\removed}[1]{\textcolor{red}{\sout{#1}}}
\renewcommand{\added}[1]{\textcolor{black}{#1}}
\renewcommand{\removed}[1]{\if0{#1}\fi}

\begin{document}

\title{Reduction of Forgetting by Contextual Variation During Encoding Using 360-Degree Video-Based Immersive Virtual Environments}


\author{
    Takato Mizuho,
    Takuji Narumi,
    and Hideaki Kuzuoka
    \thanks{Takato Mizuho, Takuji Narumi, and Hideaki Kuzuoka are with the Graduate School of Information Science and Technology, The University of Tokyo, Tokyo, Japan.}
    }

\markboth{IEEE Transactions on Visualization and Computer Graphics,~Vol.~00, No.~0, August~2023}%
{Mizuho \MakeLowercase{\textit{et al.}}: Reduction of Forgetting by Contextual Variation During Encoding Using 360-Degree Video-Based Immersive Virtual Environments}

\IEEEpubid{0000--0000/00\$00.00~\copyright~2021 IEEE}



\maketitle

\begin{abstract}
    Recall impairment in a different environmental context from learning is called context-dependent forgetting. Two learning methods have been proposed to prevent context-dependent forgetting: reinstatement and decontextualization. Reinstatement matches the environmental context between learning and retrieval, whereas decontextualization involves repeated learning in various environmental contexts and eliminates the context dependency of memory. Conventionally, these methods have been validated by switching between physical rooms. However, in this study, we use immersive virtual environments (IVEs) as the environmental context assisted by virtual reality (VR), which is known for its low cost and high reproducibility compared to traditional manipulation. Whereas most existing studies using VR have failed to reveal the reinstatement effect, we test its occurrence using a 360-degree video-based IVE with improved familiarity and realism instead of a computer graphics-based IVE. Furthermore, we are the first to address decontextualization using VR. Our experiment showed that repeated learning in the same constant IVE as retrieval did not significantly reduce forgetting compared to repeated learning in different constant IVEs. Conversely, repeated learning in various IVEs significantly reduced forgetting than repeated learning in constant IVEs. These findings contribute to the design of IVEs for VR-based applications, particularly in educational settings.
\end{abstract}


\section{Introduction}
\IEEEPARstart{W}{hen} the environmental context of learning and testing do not match, recall performance is reduced compared to when they do match~\cite{Smith-Vela2001}. This effect is referred to as context-dependent forgetting, where the environmental context refers narrowly to incidental environmental surroundings, such as places and odors, and broadly includes internal states, such as emotions and thoughts~\cite{Smith2013}.
Context-dependent forgetting is an essential topic in the field of education because it occurs unintentionally, despite one's efforts.

\begin{figure*}
\centering
\includegraphics[width=\linewidth]{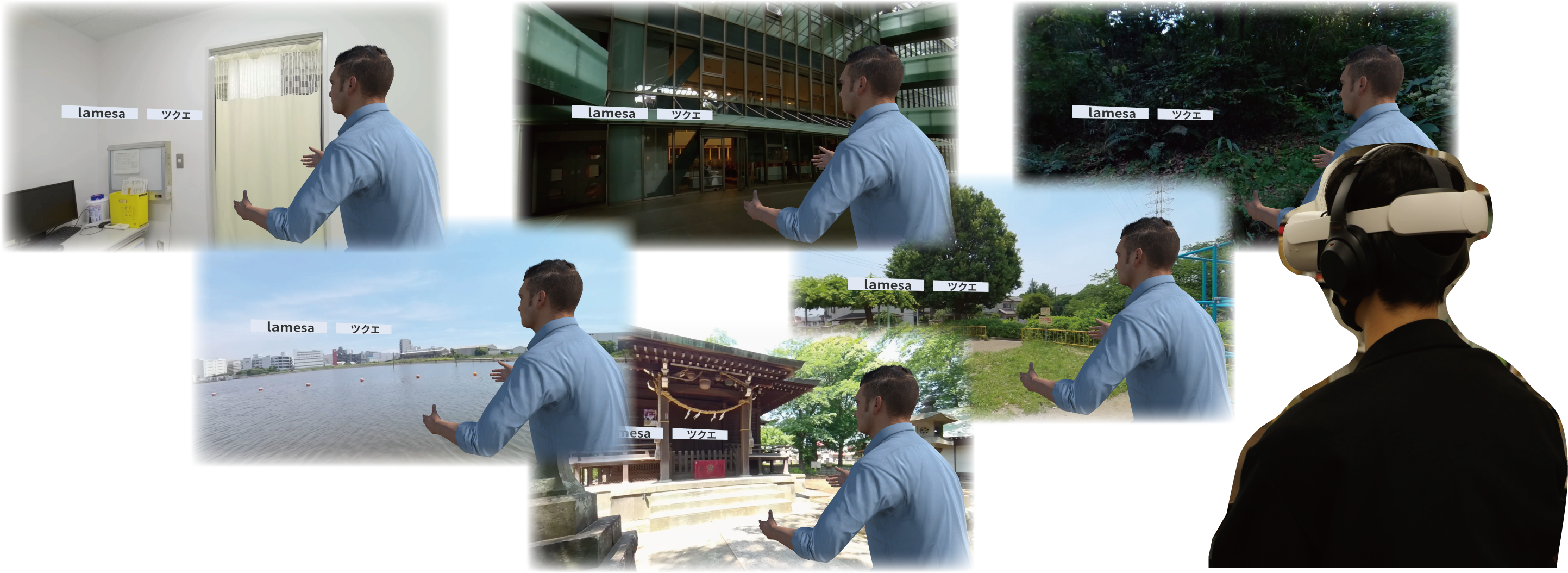}
\caption{Virtual reality (VR) offers the benefit of learning experiences in various locations, all from the comfort of our own rooms. In this study, we used VR to examine the effect of repeated learning under various environmental contexts, which made forgetting harder (decontextualization effect). In a Tagalog-Japanese pair learning task, the condition where repeated learning was conducted in diverse virtual environments significantly reduced forgetting compared to that where repeated learning was conducted in a single virtual environment.}
\label{fig:teaser}
\end{figure*}

The deleterious effect of environmental context can be alleviated by manipulating the learning contexts of people. One method is to perform advanced learning in the same context as retrieval. Godden and Baddeley~\cite{Godden-Baddeley1975} showed that when a free recall test was conducted underwater, scores were higher when the word lists were learned underwater than learned on land. This method is known as reinstatement in context-dependent memory research.

\IEEEpubidadjcol

Another method to prevent context-dependent forgetting is to learn in multiple contexts. Smith et al.~\cite{Smith-etal1978} showed that a varied condition group, who studied a word list in one room and restudied it in another room, scored higher on a final recall test performed in a room different from any room used during learning than the constant condition group, who studied and restudied the word list in the same room. This method of learning is referred to as decontextualization~\cite{Smith-Handy2014}.

When these methods are put into practice, the use of immersive virtual environments (IVEs) presented via a head-mounted display (HMD) instead of actual physical environments has been proposed~\cite{Walti-etal2019, Shin-etal2021, Watson-Gaudl2021, Rocabado-etal2022}. Using IVEs is superior to traditional physical environmental context manipulation because they do not require movement, have high reproducibility owing to electronic data representing environmental information, and have high ecological validity~\cite{Reggente-etal2018, Neo-etal2021}, such that we can measure natural cognitive activity despite being a laboratory experiment. Hence, virtual reality (VR) is suggested as a promising tool that can facilitate the conduct of experiments and accelerate the application of context-dependent memory research~\cite{Smith2013}.

However, a majority of the studies using IVEs have failed to produce context-dependent effects found in the case of physical environments. 
We posit that one reason is that all studies on IVE context have used three-dimensional computer graphics (3DCG) environments, except the study by W\"{a}lti et al.~\cite{Walti-etal2019}, who did not disclose the details of their IVEs. For example, Shin et al.~\cite{Shin-etal2021} used two IVEs on Mars and underwater.
Using 3DCG to produce context-dependent effects has two issues. First, the richness and realness of environmental information are often poor. As most 3DCG IVEs are created from scratch, they are less informative than the physical world and time- and cost-dependent. Second, the familiarity with synthesized environments is poor. Whereas Smith and Manzano~\cite{Smith-Manzano2010} selected videos of familiar scenes from everyday life to produce context-dependent effects, Shin et al.~\cite{Shin-etal2021} used artificially synthesized IVEs such as Mars with which no participants were familiar. The aforementioned characteristics may be necessary for activating hippocampal place cells, which are the neuroscientific basis for the context dependency of episodic memory~\cite{Bird-Burgess2008}.

In this study, we propose the use of 360-degree video-based IVEs. As a 360-degree video captures shots of an actual landscape, the amount of information and sense of realism are vastly improved compared to that from 3DCG. Additionally, we use 360-degree videos of familiar places. Thus, we posit that the proposed method is more likely to produce context-dependent effects than existing methods using 3DCG.

We aim to reveal (1) the reinstatement and (2) decontextualization effects using 360-degree video-based IVEs.
No study but Shin et al.~\cite{Shin-etal2021} has clarified the reinstatement effect using IVEs, but they might have overestimated it, as mentioned later. Additionally, this is the first time to study the decontextualization effect using IVEs.

We employed the experimental design adopted by Smith and Handy~\cite{Smith-Handy2014, Smith-Handy2016}, which allowed us to examine the reinstatement and decontextualization effects simultaneously.
The participants studied 20 Tagalog-Japanese pairs, performed five successive retrieval practices, and completed a final cued recall test two days later.
We primarily compared the amount of forgetting from the first day (Day 1) to the second day (Day 2) and hypothesized that learning in the same IVE on Day 1 as that on Day 2 would reduce forgetting due to the reinstatement effect and that learning in various IVEs on Day 1 would prevent forgetting due to the decontextualization effect.

Furthermore, we measured presence, which is a measure of subjective immersion in VR experience, and examined its relationship with context-dependent effects: reinstatement effect and decontextualization effects. Presence is defined as ``a sense of being there'' and is related to the formation of IVE mental models~\cite{Schubert-etal2001}. Additionally, episodic memory in VR and presence are indicated to have a close relationship~\cite{Smith-Mulligan2021}. Therefore, we considered the possibility that the higher the presence, the greater the context-dependent effects.

The present study contributed the following: 1) We proposed environmental context manipulation using 360-degree video, which is more realistic than 3DCG used by previous studies. 2) We showed a significant decontextualization effect for the first time using VR-based environmental context manipulation. 3) We achieved complicated data on the reinstatement effect, highlighting the need for further detailed investigation. 4) We measured the sense of presence, an essential indicator of VR experience, and examined its relationship with context-dependent effects for the first time.

\section{Related Work}
\subsection{Reinstatement}

The reinstatement paradigm matches the environmental context at the learning time with that at the testing time. For example, in the best-known Godden and Baddeley experiment~\cite{Godden-Baddeley1975}, participants who learned English words underwater and on land were tested underwater and on land after an interval. When participants learned underwater, they recalled better underwater than on land, whereas recall was better on land than underwater when participants leaned on land. Previous studies have used various types of environmental contexts, including place~\cite{Godden-Baddeley1975}, room~\cite{Smith1984}, odor~\cite{Isarida-etal2014}, background music~\cite{Isarida-etal2017}, and background video~\cite{Smith-Manzano2010}, which have confirmed the existence of the effect through meta-analysis~\cite{Smith-Vela2001}.

The reinstatement effect arises based on the encoding specificity principle~\cite{Tulving-Thomson1973} that episodic memories incorporate the surrounding situation or environmental context for use as a cue for later retrieval of the episode. Thus, in the case of Godden and Baddeley experiment~\cite{Godden-Baddeley1975}, the memory of words learned underwater was stored with the environmental context of being underwater; therefore, the recall was easier by reinstating the original environmental context of being underwater.

The reinstatement paradigm is simple, easy to understand, and rich in research examples. Its effectiveness is supported by meta-analysis~\cite{Smith-Vela2001}, which leads to the ease of its application as a memory support technique. However, there exist limitations such as a small effect size (\textit{d} = 0.28) suggested by the meta-analysis~\cite{Smith-Vela2001} and the need for prior information about the environmental context of testing. Hence, reinstatement is suitable for learning only the requirement in specific places and not for what needs to be demonstrated in any situation.

This study uses a 360-degree video-based IVE to test the reinstatement effect. The experimental design follows Smith and Handy~\cite{Smith-Handy2014, Smith-Handy2016}. Their experiments showed that learning progressed more quickly in the constant context condition, where the participants repeatedly learned under the constant context, than in the varied context condition, where they repeatedly learned under diverse contexts.
In addition, this study slightly modifies the experimental design to evaluate the reinstatement effect on retention and forgetting of the remembered information. Specifically, after two days of repeated learning in a constant context, we compare a condition where participants are tested in the same context with a condition where they are tested in a new context. As discussed in Section~\ref{sec:related-work:vr}, research on reinstatement effects using IVE is still in its infancy. Thus, we aim to gain new insights through this study.

\subsection{Decontextualization}

The decontextualization paradigm is used to learn in various environmental contexts. This method reduces the environmental context dependency of memory and fundamentally eliminates context-dependent forgetting. In the well-known Smith et al. experiment~\cite{Smith-etal1978}, recall performance was significantly higher when a list of words was repeatedly learned in two rooms than in the condition where the list was repeatedly learned in one room. Similar effects were confirmed not only in methods using rooms~\cite{Imundo-etal2021, Glenberg1979} but also in methods using background photos~\cite{Smith-Handy2016} and videos displayed on a monitor~\cite{Smith-Handy2014}. While learning Tagalog-English and face-name pairs, significantly less forgetting was observed in the varied condition, where the participants repeated learning with different backgrounds, than in the constant condition, where the participants repeated learning with the same background for the entire duration.

Two hypotheses have been proposed to explain the decontextualization. One hypothesis is encoding variability~\cite{Bower1972}, which states that an increase in the number of contextual cues attached to the memory to be remembered facilitates recall. For example, in the experiment by Smith et al.~\cite{Smith-etal1978}, there were twice as many contextual cues supporting retrieval under the two-room condition as under the one-room condition.
The other hypothesis is desirable difficulty~\cite{Bjork-Bjork1992}, which states that more information is encoded while retrieval is attempted when a recall is inhibited. In the varied condition, relearning occurs in contexts different from the original context, a loaded condition in which context-dependent forgetting is likely to occur, such that more information is stored in the memory.

Compared to the reinstatement effect, research on the decontextualization effect is limited, and the correctness of the hypotheses remains unclear. Contrarily, cases have reported that the superiority of the constant and varied conditions is interchanged depending on the type of environmental context being manipulated~\cite{Isarida-Isarida2010, Isarida-etal2021} or the type of repeated learning~\cite{Smith-Handy2014, Isarida-etal2021}. Considering the former, Isarida and Isarida~\cite{Isarida-Isarida2010} validated that recall was higher in the varied condition than in the constant condition when using simple-place context, where the only place was manipulated as the environmental context. Conversely, they presented that recall was better in the constant condition than in the varied condition when using complex-place context, where place, social environment (how many participants experimented at a time), and encoding task were simultaneously manipulated as the environmental context. Furthermore, considering the latter point of repeated learning type, Smith and Handy~\cite{Smith-Handy2014} showed that the use of retrieval practice (testing and feedback) produced a varied-condition-dominant decontextualization effect. Conversely, the effect disappeared and indicated a slightly constant-condition dominance when restudying (word representation only). This constant-condition dominance was replicated by Isarida et al.~\cite{Isarida-etal2021}.

Hence, the unexplored mechanisms of the decontextualization paradigm offer greater challenges than those of the reinstatement paradigm. This may be attributed to the high cost of preparing and switching between several environmental contexts, making it difficult to conduct experiments. However, the decontextualization effect is superior to the reinstatement paradigm as a memory support technique because its effect size has been suggested to be larger than that of the reinstatement effect~\cite{Smith-Vela2001}. Additionally, no prior knowledge of the environmental context during testing is required.

This study uses IVE for the first time to test the decontextualization effect. Using IVE significantly reduces the cost of preparing and moving through various contexts, among the abovementioned disadvantages. The experimental design follows Smith and Handy~\cite{Smith-Handy2014, Smith-Handy2016}, as they have consistently detected decontextualization effects.
We test the hypothesis that repeated learning under diverse IVEs would reduce forgetting and improve retention compared to using a single constant IVE.

\subsection{Environmental Context Manipulation Using Virtual Reality}
\label{sec:related-work:vr}

While previous studies have conducted experiments mainly using rooms, recent studies use VR to manipulate environmental context. For example, W\"{a}lti et al.~\cite{Walti-etal2019}, and Shin et al.~\cite{Shin-etal2021} worked on the reinstatement paradigm. In addition, Watson and Gaudl~\cite{Watson-Gaudl2021} and Rocabado et al.~\cite{Rocabado-etal2022} examined the effect of splitting a single word list into multiple sub-lists and learning each in a different context.
However, the effect was different from decontextualization, which uses multiple environmental contexts but repeatedly learns a single list.

Most existing studies that use VR-based methods have failed to identify significant context-dependent effects.
Only Shin et al.~\cite{Shin-etal2021} found the reinstatement effect using IVEs. However, Shin et al. may have overestimated the effect owing to their encoding task, where participants rated to-be-remembered items in terms of their usefulness for survival in the surrounding IVEs. Thus, the IVEs were encoded as focal information rather than incidental information and may have deviated from the definition of the environmental context~\cite{Bjork-RichardsonKlavehn1989, Smith1988}.

All of the aforementioned studies used 3DCG imaginary IVEs, except for W\"{a}lti et al.~\cite{Walti-etal2019} who did not disclose the information related to their IVEs in detail. In contrast, previous traditional studies used real places and their pictures or videos. It remains unclear whether imaginary landscapes are as effective as real ones, which may be the reason why IVEs could not have consistently produced context-dependent effects. From the neuroscientific perspective, whether the stimuli led the hippocampal place cells to fire is crucial to produce context-dependent effects~\cite{Bird-Burgess2008}. However, to the best of our knowledge, it is not yet clear what kind of virtual environment would cause place cells to fire as in reality~\cite{Reggente-etal2018}. In this study, we believe that 360-degree videos of actual places are more suitable as stimuli for producing environmental context-dependent effects than imaginary 3DCG.
Given that two-dimensional live-action videos and photos can induce context-dependent effects~\cite{Smith-Manzano2010, Smith-Handy2014}, we posit that 360-degree video-based IVEs, which extend them in all directions, would also produce context-dependent effects.
Therefore, we propose environmental context manipulation using live-action 360-degree videos.

Additionally, previous studies on context-dependent effects using VR have not fully examined important indicators of the VR experience. The most important measure is presence. Presence is the sense of being in an IVE~\cite{Schubert-etal2001}. 
The degree of presence is known to affect several aspects of cognition and behavior in IVE.
Specifically, Smith reviewed studies of episodic memory using VR and suggested the close relationship between episodic memory and presence~\cite{Smith2019}. This relationship may be based on the fact that presence and episodic memory are related to spatial recognition and depend on the functioning of the hippocampus~\cite{Bird-Burgess2008, Cummings-Bailenson2016}. Context-dependent effects are one of the phenomena of episodic memory, which is likely to be affected by presence. However, this has not been investigated in previous studies. Therefore, this study measures presence using questionnaires and discusses its relationship with context-dependent effects.
\added{Although we employed a widely used questionnaire to measure presence in VR research, it should be noted that there is a debate as to whether such a questionnaire can truly measure presence~\mbox{\cite{Slater2004}}.}

Smith also suggests that memory research using VR should consider the physical discomfort, or VR sickness, that occurs during VR experiences~\cite{Smith2019}. VR sickness is caused by discrepancies between visual and vestibular sensations~\cite{Palmisano-etal2017} and by prolonged VR experiences~\cite{Kennedy-etal2000}. In this experiment, participants were always seated and did not move around within the IVE. Thus, it was unlikely that severe VR sickness would occur. However, it should be noted that the total length of the VR experience was long. Therefore, VR sickness was measured with a questionnaire to examine if it was a confounding factor that interfered with memory.

This study examines two methods for preventing context-dependent forgetting, reinstatement and decontextualization. This study is the first to investigate decontextualization effects using VR.
While previous studies have used 3DCG-based IVEs, this study uses 360-degree video-based IVEs. In addition, we consider indicators specific to VR experiences, such as presence and VR sickness, which have not been considered in previous studies, and discuss their relationship to the observed memory phenomena.

\section{Experiment}
\subsection{Participants}
\label{participants}
Sixty-three healthy native Japanese speakers (fifty-four males and nine females aged 26.4 $\pm$ 1.1 (\textit{SE}) years) participated in the experiment.
The sample size was determined as follows: We conducted a preliminary experiment (\textit{N} = 9) to ensure the validity of the experimental system and method. Consequently, the effect size for the forgetting score, which was the main indicator as mentioned in the later section, was \textit{f} = 0.47. When designing the sample size using G*Power~\cite{Faul-etal2007} using this effect size, it was calculated that 48 participants were needed (ANOVA:Fixed effects, omnibus, one-way; $\alpha$ = 0.05; Power ($1 - \beta$) = 0.8; Number of groups = 3). As it was a three-condition between-participants design (Section~\ref{sec:design}), the sample size should be a multiple of three. In addition, considering the counterbalance of the IVE used (Section~\ref{sec:context}), the sample size needed to be a multiple of seven. Consequently, the total sample size had to be a multiple of 21. Therefore, the final sample size was set at 63.

As the experiment was conducted remotely, we recruited only those who possessed the equipment required for the experiment described later, such as an HMD.
We asked the participants about their previous VR experience using a five-point Likert scale, ranging from ``never (0)'' to ``use very often (4).'' The overall average score was 2.87 $\pm$ 0.1 (\textit{SE}).
The participants were paid after the second day of the experiment.
The experimental protocol was approved by the local ethics research committee of the Graduate School of Information Science and Technology, The University of Tokyo (UT-IST-RE-210806).

\begin{figure*}[tb]
\centering
\includegraphics[width=\linewidth]{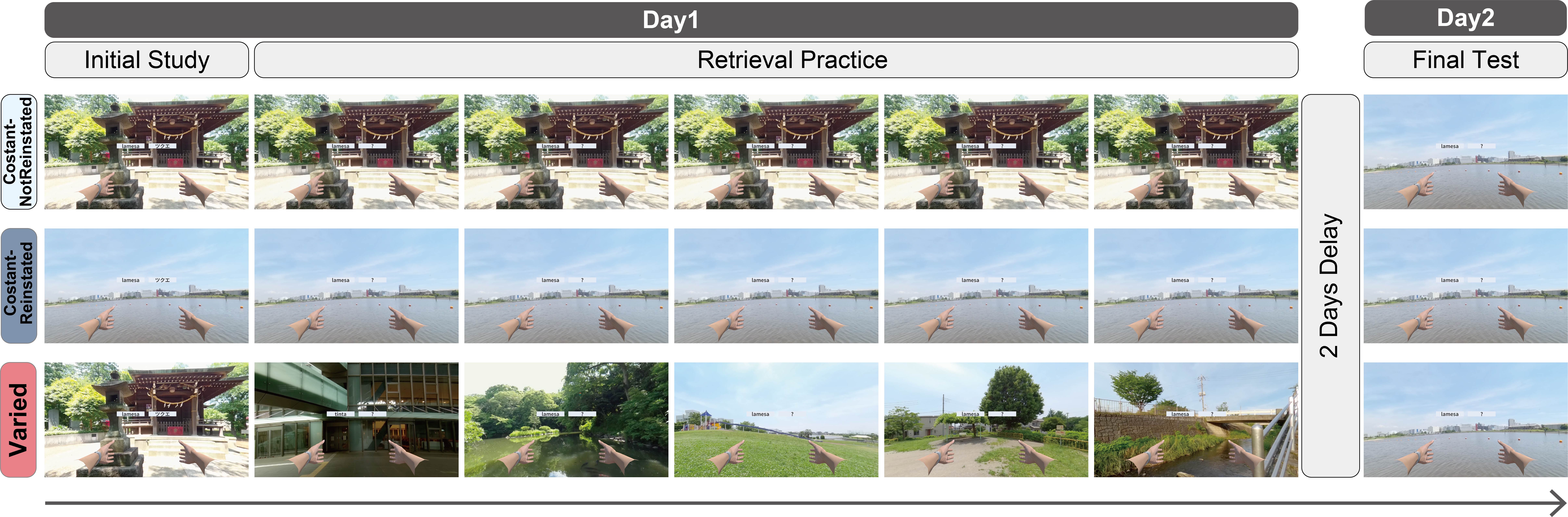}
\caption{Experimental design. All participants studied 20 Tagalog-Japanese pairs, took five retrieval practices, and had a final test two days later. In the Constant-NotReinstated condition, a participant performed an initial study, five retrieval practices in a constant IVE, and a final test in a new IVE.
In the Constant-Reinstated condition, a participant performed an initial study, five retrieval practices in a constant IVE, and a final test in the same IVE.
In the Varied condition, a participant performed an initial study and five retrieval practices in IVEs different from each other, and then a final test was conducted in a new IVE. We prepared seven IVEs and counterbalanced them among the participants. The number of words correctly recalled on the five retrieval practices and final test and forgetting from Day 1 to Day 2 were measured and compared between participants in the three conditions.}
\label{fig:design}
\end{figure*}

\subsection{Experimental Design}
\label{sec:design}
As shown in \cref{fig:design}, the experiment was conducted in a three-condition between-participants design; Constant-NotReinstated, Constant-Reinstated, and Varied conditions. In the Constant-NotReinstated condition, a participant performed an initial study, five retrieval practices in a constant IVE, and a final test in a new IVE. In the Constant-Reinstated condition, a participant performed an initial study, five retrieval practices in a constant IVE, and a final test in the same IVE. Whereas in the Varied condition, a participant performed an initial study and five retrieval practices in an IVE different from each other, and then a final test was conducted in a new IVE. We prepared seven IVEs and counter-balanced them among the participants. More details on IVEs are disclosed in later sections.

Participants were randomly assigned to each condition. However, we considered the gender balance, age, and degree of VR experience to be comparable. The Constant-NotReinstated group comprised 18 males and 3 females, with an average age of 25.8 $\pm$ 1.6 (\textit{SE}). The Constant-Reinstated group also consisted of 18 males and 3 females, with an average age of 26.7 $\pm$ 2.3 (\textit{SE}). Similarly, the Varied group included 18 males and 3 females, with an average age of 26.7 $\pm$ 1.8 (\textit{SE}). Regarding the mean level of previous VR experience, the Constant-NotReinstated group scored 2.67 $\pm$ 0.16 (\textit{SE}), the Constant-Reinstated group scored 2.90 $\pm$ 0.22 (\textit{SE}), and the Varied group scored 3.05 $\pm$ 0.19 (\textit{SE}).

\subsection{Apparatus}
The physical settings of the experimental setup are shown in \cref{fig:teaser}.
The participants experimented in a seated position as
they wore an HMD, headphones or earphones, and two controllers for the VR experience.
Additionally, the participants used their personal computers or smartphones to answer Google form questionnaires before, between, and after the VR experiences.

\begin{figure}[tb]
    \centering
    \subfloat[First-person perspective during learning.]{\includegraphics[width=\linewidth]{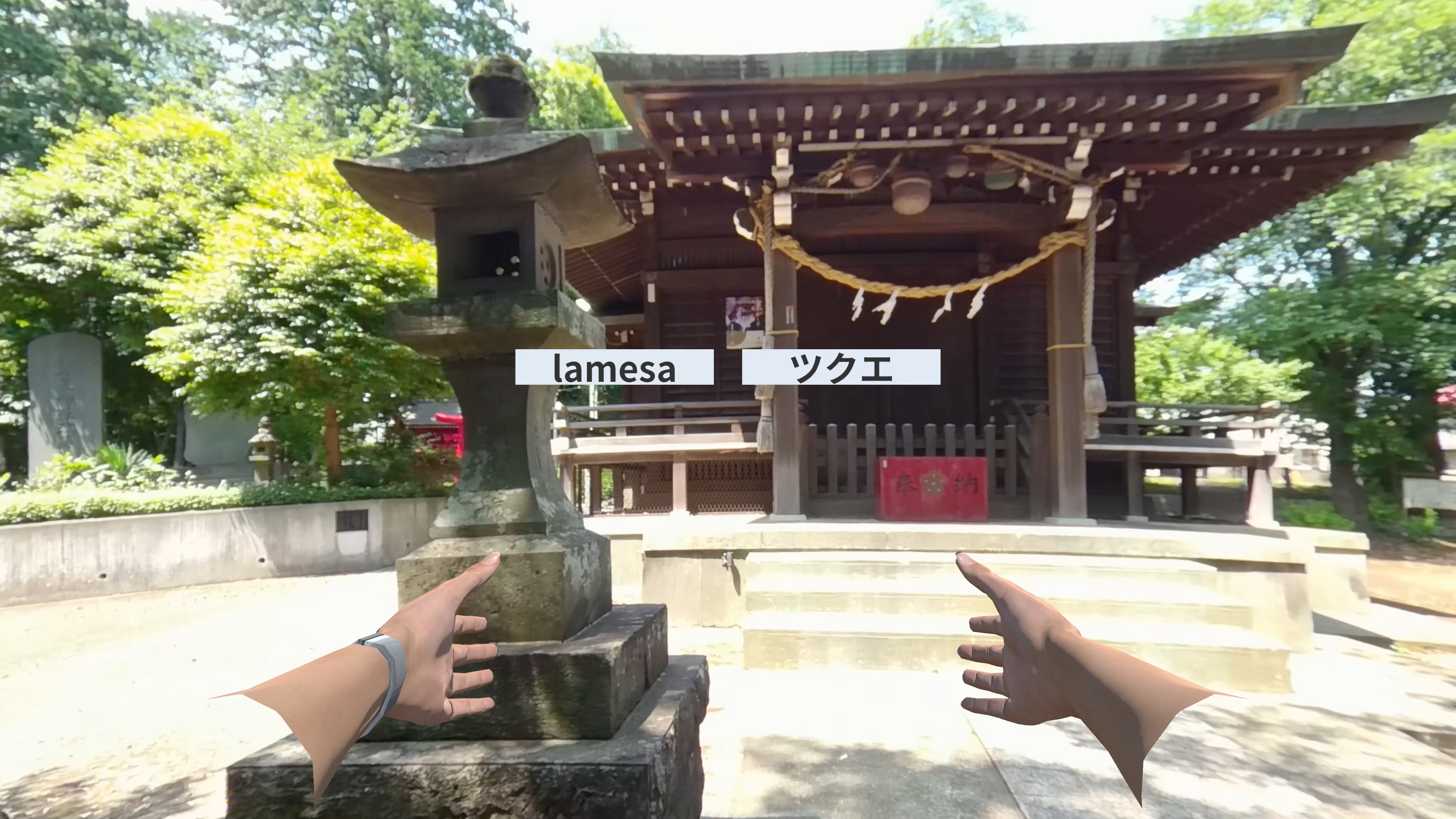}%
    \label{fig:system:learn}}
    
    \subfloat[First-person perspective during testing.]{\includegraphics[width=\linewidth]{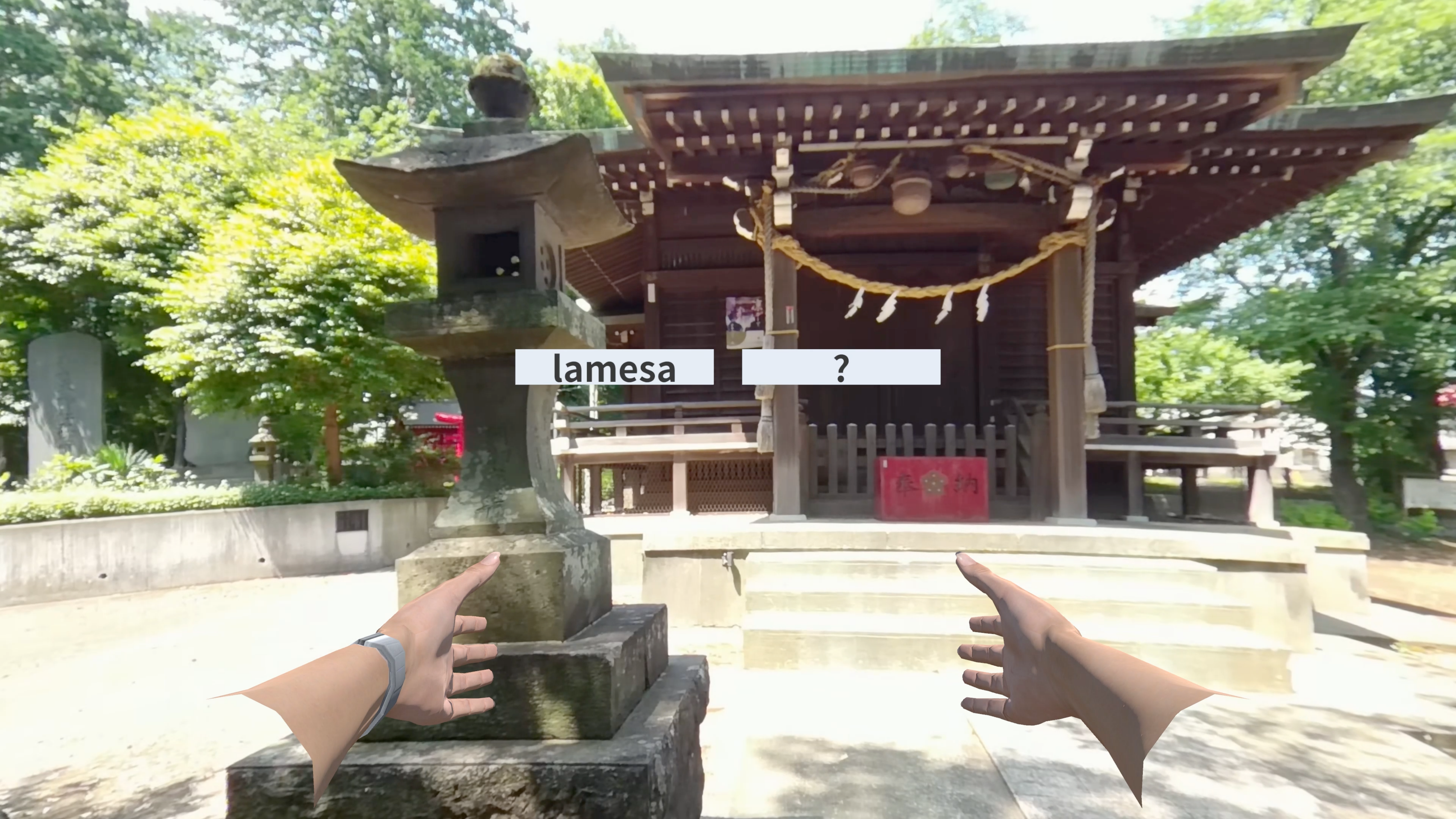}%
    \label{fig:system:test}}
    \caption{Participants engaged in a memory task to memorize 20 Tagalog-Japanese pairs of words. (a) During learning, the Tagalog words were presented in the left box in front, while their Japanese translations were presented in the right box at a rate of 5 s/pair. (b) Only the Tagalog words were presented during testing, with a question mark in the right box. The participants responded orally with the Japanese translation corresponding to the Tagalog word.}
    \label{fig:system}
\end{figure}

The IVEs were rendered by Unity\footnote{\url{https://unity.com/}} (Version 2019.4.9f1) and presented via Meta Quest2 HMD\footnote{\url{https://www.meta.com/quest/products/quest-2/}} with a resolution of 1832 $\times$ 1920 per eye and a refresh rate of 90 Hz.
The participants held two Meta Touch controllers and interacted with the IVEs.
The surrounding scenery and environmental sounds were manipulated according to the assigned condition. Furthermore, the participants were able to listen to instruction voice via headphones or earphones, such as an instruction for starting the recall test.
The instruction voice was a female voice created by Amazon Polly\footnote{\url{https://aws.amazon.com/polly/}}.

The participants used avatars to represent their bodies virtually in the IVEs. We selected two avatars from the Microsoft Rocketbox Avatar Library~\cite{GonzalezFranco-etal2020}, one for male (Male\_Adult\_08) and the other for female (Female\_Adult\_01) participants. To eliminate the possibility of gender in-congruence affecting cognition~\cite{Schwind-etal2017}, participants used an avatar of the same gender as their own.
Using the library, which is open source, helps improve the reproducibility of research. We selected the avatars as neutral as possible to avoid confoundings, with no specific meaning or appearance. As the participants were Asian, an Asian avatar might have been a better choice. However, we could not select an appropriate Asian avatar from the library. We did not consider this a serious problem because it has been suggested that users do not significantly lose their sense of embodiment to avatars of a different ethnicity than themselves~\cite{Marini-Casile2022}.
The full body movements of the avatar were controlled based on the inverse kinematic calculations performed by the Root Motion Final IK software\footnote{\url{https://assetstore.unity.com/packages/tools/animation/final-ik-14290}} utilizing the information on the positions and rotations of the HMD and two hand-held controllers.

\subsection{To-be-remembered Items}
Twenty Tagalog words with Japanese translations were selected as follows. First, three-letter Japanese nouns with familiarities between 6.00 and 6.99 were chosen~\cite{Koyanagi-etal1960English} because familiarity is an indicator related to memorability. Second, the chosen nouns were translated into Tagalog using Google Translation and Glosbe\footnote{\url{https://ja.glosbe.com/}}, free translational services available on the Internet. Third, the translated Tagalog words were retranslated into Japanese, and any words incongruently translated were removed. Finally, 20 words of two to four syllables were selected, provided none were obvious English cognates, because word length is another measure related to memorability. These selection criteria followed that of a previous study~\cite{Smith-Handy2016}.

As shown in \cref{fig:system}, the items were visually presented five meters in front of the participant in black letters on a white background, with Tagalog on the left and Japanese on the right side.
In the recall tests, only Tagalog words were presented with ``?'' on the right instead of the Japanese words. The height of these letters was adjusted to be at eye level. Their font size was also adjusted so that the vertical field of view was approximately three degrees.

\subsection{Environmental Contexts}
\label{sec:context}
\begin{figure*}[tb]
\centering
\includegraphics[width=\linewidth]{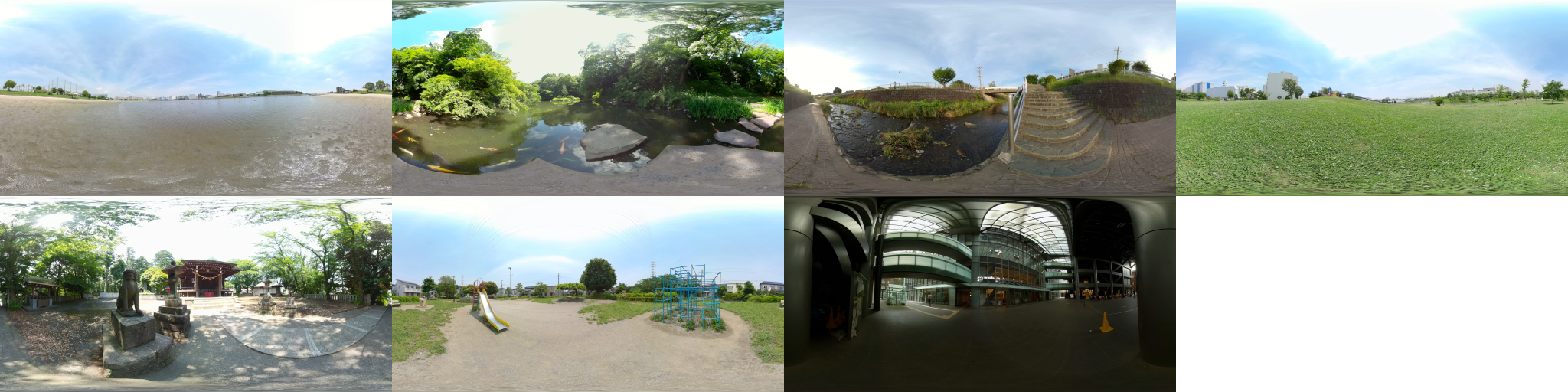}
\caption{Shots of 360-degree videos used as environmental context. From top left: Beach, Pond, River, Glassland, Shrine, Park, and Building. The shots were taken by a KANDAO Qoocam8K camera with a resolution of 7680 $\times$ 3840, 30 fps, and monoscopic.}\label{fig:context}
\end{figure*}

Seven 360-degree videos, as shown in \cref{fig:context}, were used as IVEs. The videos were taken using KANDAO Qoocam8K camera\footnote{\url{https://www.kandaovr.com/qoocam-8k/index.html}} with a resolution of 7680 $\times$ 3840, 30 fps, and monoscopic. These videos were selected to be perceptually and semantically distinct from each other. Thus, their contents included a beach, pond, river, shrine, park, building, and grasslands.

In the Constant-NotReinstated condition, two IVEs were randomly selected, one for the initial study and the retrieval practices and another for the final test. However, we balanced IVEs among the participants such that each IVE was used the same number of times for the final test. In the Constant-Reinstated condition, one of the IVEs was randomly selected for the initial study, retrieval practices, and final test. However, the choice of IVEs for use was balanced among the participants. In the Varied condition, each of the seven IVEs was used once, and the order in which they were used was randomized. The IVEs were balanced among the participants such that each IVE was used the same number of times for the final test.

\subsection{Procedures}
The participants experimented remotely from their homes (or other private places) after reading the instructions distributed to them; they experimented on their accord. The video recordings from the HMDs and the execution recordings of the experimental application were used to verify the adequateness of the experiments performed by the participants. The experiment lasted for a total of two days.

\subsubsection{Day 1}
Day 1 began with an adaptation section. First, the participants answered a Google form about their physical condition at the start of the experiment, their bedtime, and their waking time on the previous day. Thereafter, they wore the HMD and headphones (or earphones) and observed the presented IVE for one minute. The HMD was then removed, and the participants were to answer a Japanese version of the igroup presence questionnaire (IPQ)~\cite{Schubert-etal2001} to measure their sense of presence. The IPQ consisted of 14 questions, which the participants were to answer, each on a seven-point scale from one to seven. Each item had a unique anchor at each end of the scale, with one corresponding to low presence and seven to high presence, except for three reversed items. After reversing the scores on the inverted items, for instance, reversing two to six, the average of all items was calculated as the presence score. This adaptation protocol was performed once for each of the seven IVEs.

An initial study section followed the adaptation section. After re-wearing the HMD again and observing the IVE for 30 seconds, the participants were presented with 20 consecutive Tagalog-Japanese pairs of words at a rate of five seconds/pair and instructed to memorize them. The order of the presentation was randomized among the participants. The participants were informed in advance about the test that would be conducted, where they were required to respond with the Japanese translation when presented with the Tagalog words. After the presentation, the participants were asked to remove their HMD.

Next, the participants had a retrieval practice section, where they wore their HMD and observed the IVE for 30 seconds, and then conducted a retrieval practice. The retrieval practice proceeded as follows. First, a Tagalog word was presented, to which the participants responded orally with the corresponding Japanese translation. The Tagalog word remained on the screen for five seconds, during which the participants had to respond. Immediately after the response time was over, the correct Tagalog-Japanese pair was presented and remained on the screen for five seconds, given as feedback. During this feedback period, the participants completed a self-score using two floating cubes at hand. When the response was correct, the participants touched the cube marked as ``O,'' while for an incorrect response, they touched the cube marked as ``X,'' using the hands of their avatar. The participants repeated the above procedure for all 20 pairs of words and then removed their HMD.

The retrieval practice section was performed five times in succession. The order of presentation for each pair was randomized among the five retrieval practices.
Besides removing the HMD, no explicit breaks were provided between the retrieval practice sections. The participants were allowed to proceed at their own pace but were instructed to complete the experiment in batches unless they felt VR sickness.
Thereafter, the participants completed a simulator sickness questionnaire (SSQ)~\cite{Kennedy-etal1993}, which is used to measure VR sickness. With this, the first day of the experiment came to an end. The SSQ consisted of 16 questions, answered on a four-point scale of none (0), slight (1), moderate (2), and severe (3). The scores for each item were summed to give scores for nausea, oculomotor, and disorientation, and finally, the total score was calculated from the three sub-scales representing the overall severity.

\subsubsection{Day 2}
Two days after the last retrieval practice (approximately 47 h later), the second day of the experiment started. First, the participants were asked to respond to a Google form about their physical condition at the start of the experiment and their bedtime and wake-up time on the previous day. Next, the participants were asked to wear their HMD and observe the IVE for 30 s. Thereafter, a cued recall test was conducted as the final test. In the cued recall test, a Tagalog word was presented, and the participants were to orally answer with the corresponding Japanese translation within the response time of 5 s per pair. The test was repeated consecutively for all the 20 pairs of words, whose order of presentation was randomized. Finally, the participants removed their HMD and answered the IPQ, SSQ, and questionnaire about the whole experiment. With this, the second day of the experiment came to an end.

\section{Measurements and Hypotheses}
We reviewed the collected recordings and scored the oral responses for the five retrieval practices and final test.
Thereafter, four indicators following Smith and Handy were analyzed~\cite{Smith-Handy2014}, which included first retrieval practice (RP1), acquisition, retention, and forgetting. RP1 represented the number of correct answers achieved in the first retrieval practice. During the first retrieval practice, the Constant-NotReinstated and Constant-Reinstated conditions reinstated the original environmental context, whereas the Varied condition did not. Therefore, based on the reinstatement effect, we hypothesized that the Constant-NotReinstated and Constant-Reinstated conditions were superior to the Varied condition.

Acquisition defined the transition of the number of correct answers attained in five retrieval practices. During the five retrieval practices, the Constant-NotReinstated and Constant-Reinstated conditions reinstated the original environmental context in every retrieval practice. Conversely, the Varied condition varied the context of every retrieval practice. Therefore, based on the reinstatement effect, we hypothesized that the Constant-NotReinstated and Constant-Reinstated conditions were superior to the Varied condition for acquisition.

Retention represented the number of correct answers obtained in the final test on Day 2. During the final test, the Constant-NotReinstated condition did not reinstate the original context. Hence, context-dependent forgetting was expected to occur. Conversely, in the Constant-Reinstated condition, the original context was reinstated in the visual and auditory senses. However, the other senses, including internal states, changed temporally from Day 1 to Day 2. Hence, context-dependent forgetting was assumed to be smaller here than in the Constant-NotReinstated condition. Furthermore, we hypothesized that no context-dependent forgetting had occurred in the Varied condition due to the decontextualization effect, although the original environmental context was not reinstated.

Forgetting was the amount of information forgotten from Day 1 to Day 2.
We assessed forgetting in two ways, which followed that of Smith and Handy~\cite{Smith-Handy2014}. First, we calculated a forgetting score by subtracting the number of correct answers obtained in the final test from those obtained in the fifth retrieval practice. Second, we calculated a forgetting ratio by dividing the forgetting score by the number of correct answers obtained in the fifth retrieval practice.
Thus, describing the number of correct answers on the fifth retrieval practice as RP5 and those on the final test as FT, forgetting score and ratio can be calculated as follows.
\[
Forgetting~Score = RP5 - FT
\]
\[
Forgetting~Ratio = \frac{RP5 - FT}{RP5}
\]
Similar to that for retention, we hypothesized that forgetting was the highest in the Constant-NotReinstated condition, followed by the Constant-Reinstated condition, and hardly any occurrence in the Varied condition.
Retention and forgetting may appear to be the same metric. However, Smith and Handy suggested that forgetting is more appropriate for detecting decontextualization effects because the amount originally remembered, i.e., RP5, can be considered in the analysis.

Moreover, we measured the presence of IVEs in the experiment using the IPQ. As all IVEs were of the same quality, no differences between IVEs and between conditions were expected. Thus, we calculated the average value of the presence of one participant to seven different IVEs as the representative value and used it to investigate the relationship between presence and context-dependent effects. Specifically, we divided participants into high- and low-presence groups and compared forgetting across conditions by following Rupp et al.~\cite{Rupp-etal2016}. We hypothesized that the reinstatement and decontextualization effects were larger in the high-presence group because higher presence indicated a greater likelihood of cognitive activity equivalent to having been there.

\section{Results}
\subsection{RP1}
\begin{figure}[tb]
\centering
\includegraphics[width=\linewidth]{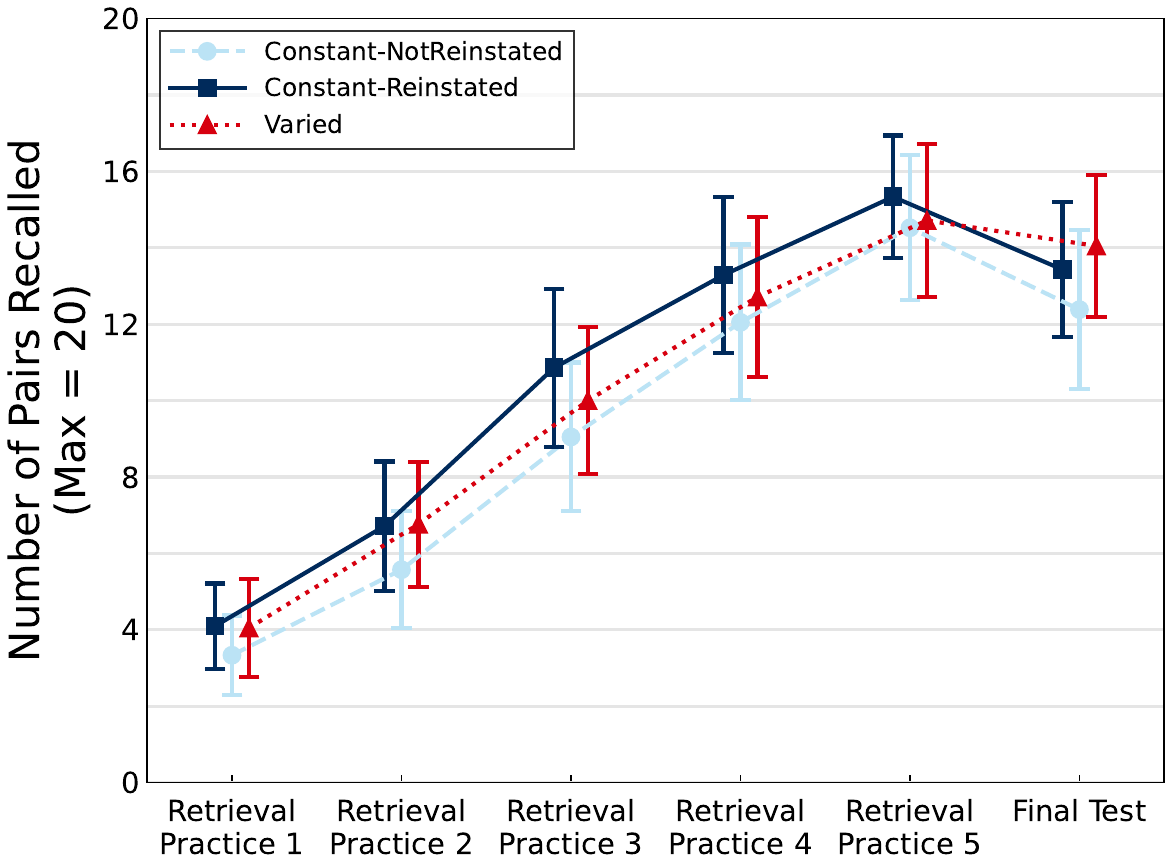}
\caption{Mean number of correct answers obtained in the five retrieval practices and the final test. Error bars represent the 95\% confidence intervals. Three ANOVA for the number of correct responses obtained in the first retrieval practice (RP1), the change in the number of correct responses during the five retrieval practices (acquisition), and the number of correct responses in the final test (retention) showed no significant differences between the conditions.}\label{fig:recall}
\end{figure}

The number of word pairs correctly recalled during the five retrieval practices and the final test is presented in \cref{fig:recall}. For evaluating RP1, a Kruskal-Wallis test showed that the main effect was not significant ($\chi^2$ (2) = 1.08, \textit{p} = .583). 

\subsection{Acquisition}

After aligned rank transformation~\cite{Wobbrock-etal2011}, which enabled us to conduct an analysis of variance (ANOVA) on non-parametric data, three by five mixed ANOVA was computed, with context (Constant-NotReinstated, Constant-Reinstated, and Varied) as a between-participants variable, retrieval practice trial (RP1, RP2, RP3, RP4, and RP5) as a repeated measure, and the number of words correctly recalled on each retrieval practice trial as the dependent variable. The results showed that only the main effect of the retrieval practice trial was significant (\textit{F} (4, 240) = 388.3, \textit{p} $<$ .001, ${\eta_p}^2$ = .87). Conversely, the main effect of context and the interaction were not significant (\textit{F} (2, 60) = 0.59, \textit{p} = .56, ${\eta_p}^2$ = .02; \textit{F} (8, 240) = 0.30, \textit{p} = .96, ${\eta_p}^2$ = .01, respectively).

\subsection{Retention}
The data with the angular transformation applied to the proportion of correct responses at the final cued recall test were used for retention analysis. As normality and homogeneity of variance were not violated (Shapiro-Wilk test, \textit{p} = .22; Levene test, \textit{p} = .66), one-way ANOVA was computed to show that the main effect was not significant and the effect size was small (\textit{F} (2, 60) = 0.75, \textit{p} = .48, ${\eta_p}^2$ = 0.02).

\subsection{Forgetting}
\label{result:forgetting}
\begin{figure*}
    \centering
    \subfloat[Forgetting score]{\includegraphics[width=0.4\linewidth]{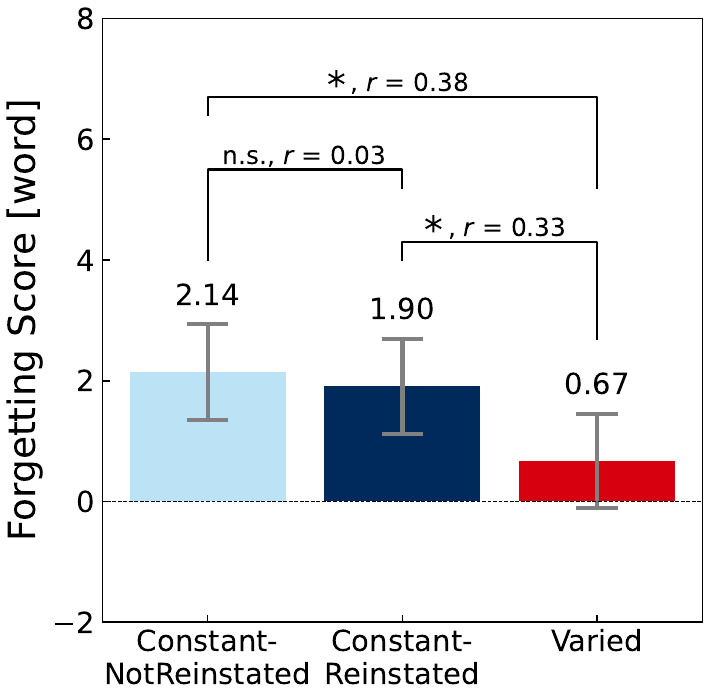}
    \label{fig:forgetting_diff}}
    \hfil
    \subfloat[Forgetting ratio]{\includegraphics[width=0.4\linewidth]{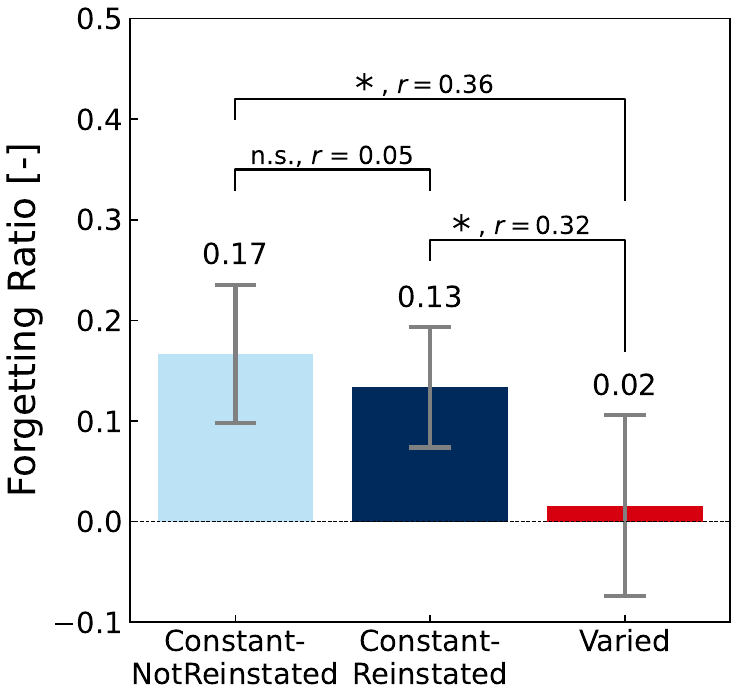}
    \label{fig:forgetting_ratio}}
    \caption{Forgetting from Day 1 to Day 2. Error bars represent the 95\% confidence intervals. (a) the number of forgotten words. (b) the ratio of forgotten words. For both measures, forgetting was significantly lower in the Varied condition than in the other two conditions, confirming the decontextualization effect. $^{*}$: \textit{p} $<$ .05.}
    \label{fig:forgetting}
\end{figure*}

The scores for forgetting are presented in \cref{fig:forgetting_diff}. A Kruskal-Wallis test showed that the main effect was significant ($\chi^2$ (2) = 6.98, \textit{p} = .030). As the main effect was significant, we repeated three Wilcoxon's rank sum tests under Shaffer's modified sequentially rejective Bonferroni method. We showed that the score for forgetting in the Varied condition was significantly less than that in both the Constant-NotReinstated and Constant-Reinstated conditions (\textit{W} = 315.0, \textit{p} = .014, \textit{r} = 0.38; \textit{W} = 303.0, \textit{p} = .033, \textit{r} = 0.33). No significant difference between the scores obtained in the Constant-NotReinstated and Constant-Reinstated conditions was observed (\textit{W} = 211.5, \textit{p} = .823, \textit{r} = 0.03).

The forgetting ratios are presented in \cref{fig:forgetting_ratio}. As a Kruskal-Wallis test showed that the main effect was significant ($\chi^2$ (2) = 6.57, \textit{p} = .037), we repeated three Wilcoxon's rank sum tests under Shaffer's modified sequentially rejective Bonferroni method. We showed that the forgetting ratio was significantly less in the Varied condition than in both the Constant-NotReinstated and Constant-Reinstated conditions (\textit{W} = 312.0, \textit{p} = .020, \textit{r} = .36; \textit{W} = 303.5, \textit{p} = .036, \textit{r} = 0.32). No significant difference between the results of the Constant-NotReinstated and Constant-Reinstated conditions was observed (\textit{W} = 208.0, \textit{p} = .760, \textit{r} = 0.05).

\subsection{Presence}
\begin{figure}
    \centering
    \subfloat[IPQ scores for the seven IVEs.]{\includegraphics[width=0.45\linewidth]{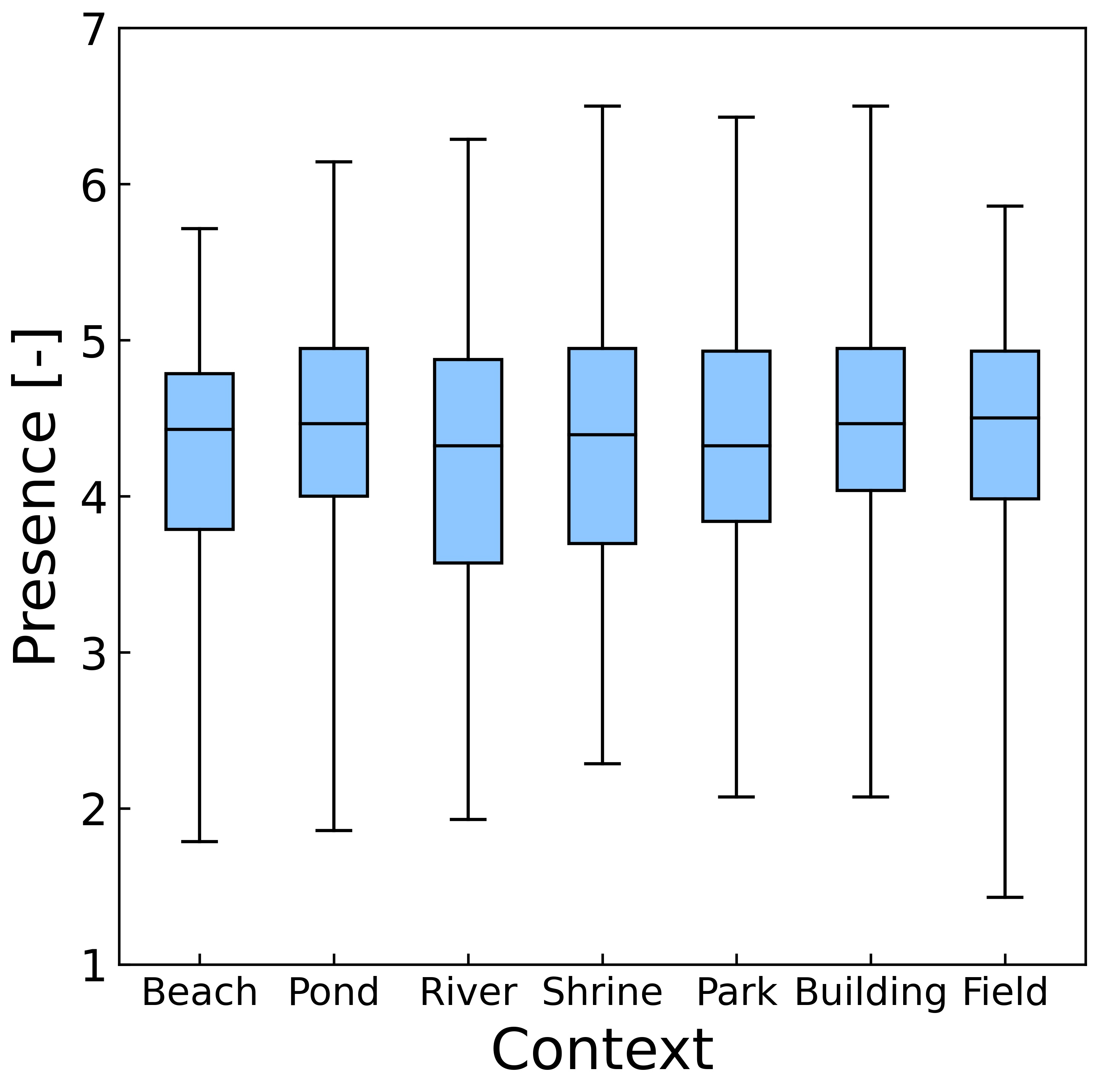}
    \label{fig:presence:context}}
    \hfil
    \subfloat[IPQ scores across the three context conditions.]{\includegraphics[width=0.45\linewidth]{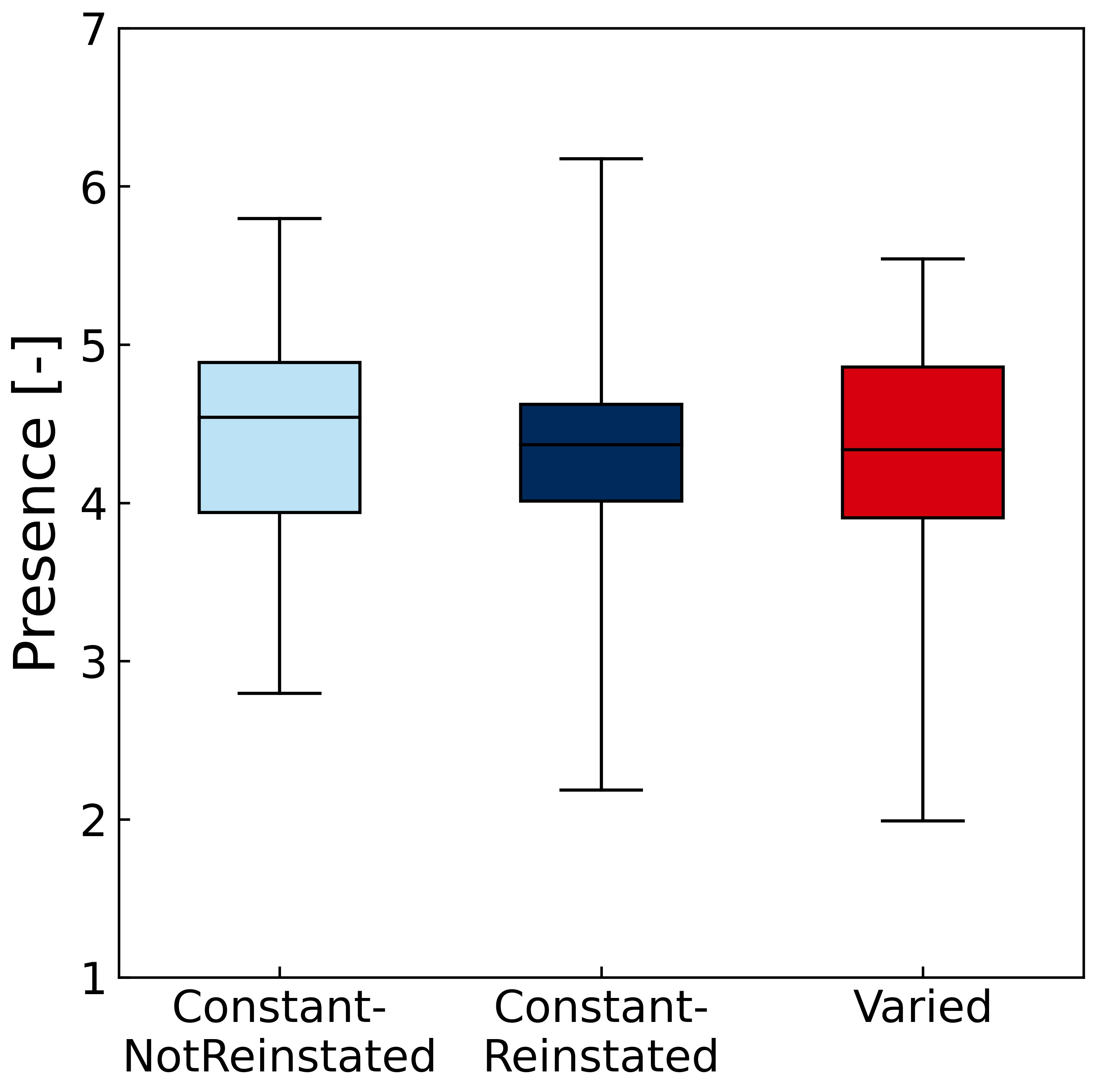}
    \label{fig:presence:condition}}
    \caption{Left panel shows the IPQ scores, or presence, for the seven IVEs. No significant differences between the IVEs were observed. The right panel shows the IPQ scores averaged for seven IVEs, considering the representative of each participant's presence. No significant differences were observed between the three conditions.}
    \label{fig:presence}
\end{figure}

During the adaptation section of Day 1, the participants answered IPQs once for each of the seven IVEs. List-wise, we removed the data of seven participants owing to missing information. The IPQ scores for the seven IVEs are presented in \cref{fig:presence:context}. A Friedmann test showed no significant difference between the IVEs ($\chi^2$ (6) = 6.61, \textit{p} = .36).

An average of the scores obtained for the seven IVEs was calculated as a representative value of the sense of presence each participant perceived. In this analysis, the mean values for six of the seven aforementioned participants were calculated, excluding missing data, and included in the analysis. The remaining participant in the Varied condition was excluded from the analysis because they did not have any of the scores. The IPQ scores averaged over the scores for seven IVEs are presented in \cref{fig:presence:condition}. A Kruskal-Wallis test showed no significant difference between the three conditions ($\chi^2$ (2) = 0.78, \textit{p} = .68).



With reference to Rupp et al.~\cite{Rupp-etal2016}, we divided all participants into two groups using the representative score of presence: those with high presence and those with low presence.
In the high-presence group, there were 13 participants were present in the Constant-NotReinstated condition, 9 in the Constant-Reinstated condition, and 9 in the Varied condition.
In the low-presence group, there were 8 participants were present in the Constant-NotReinstated condition, 12 in the Constant-Reinstated condition, and 11 in the Varied condition.

For each presence group, we compared the reinstatement effect, such as the difference in the amount of forgetting between the Constant-NotReinstated and Constant-Reinstated conditions, and the decontextualization effect, such as the difference in the amount of forgetting between the Constant-NotReinstated and Varied conditions. Thus, we performed four Wilcoxon's rank sum tests using the Holm-Bonferroni method.

\begin{figure*}
    \centering
    \subfloat[Forgetting score.]{\includegraphics[width=0.4\linewidth]{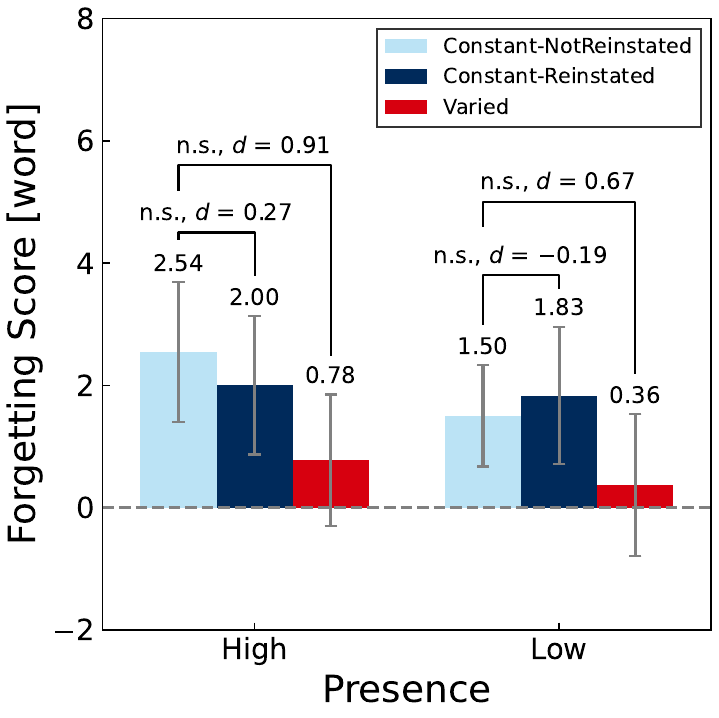}
    \label{fig:forgetting_score-presence}}
    \hfil
    \subfloat[Forgetting ratio.]{\includegraphics[width=0.4\linewidth]{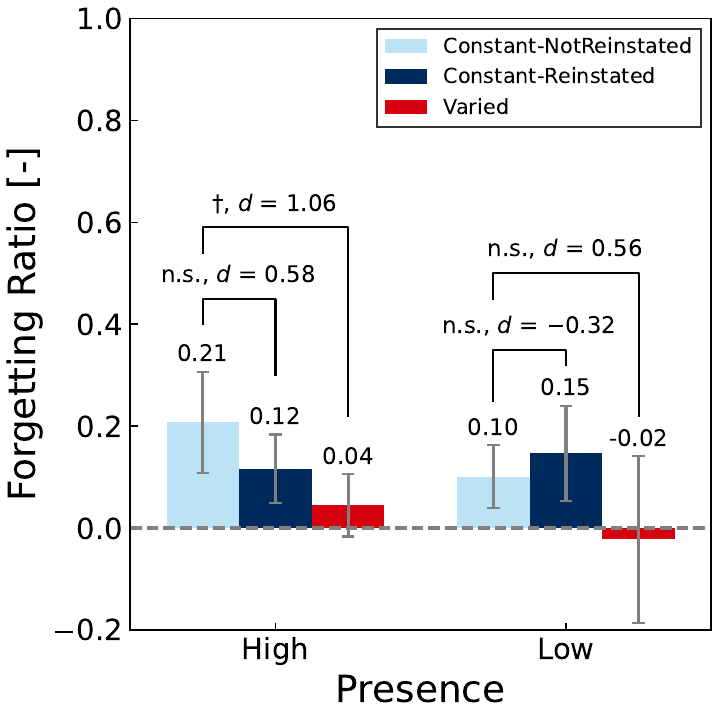}
    \label{fig:forgetting_ratio-presence}}
    \caption{Interaction between presence and forgetting. Error bars represent the 95\% confidence intervals. The left panel shows the number of forgotten words, while the right panel shows the ratio of forgotten words. For the reinstatement effect (Constant-NotReinstated vs. Constant-Reinstated), the superiority of the Constant-NotReinstated and Constant-Reinstated conditions was reversed in the high- and low-presence groups. However, the observation was not significant.}
    \label{fig:interaction-presence-forgetting}
\end{figure*}

The forgetting scores for each presence group are presented in \cref{fig:forgetting_score-presence}. In the high-presence group, no significant difference between the Constant-NotReinstated and Constant-Reinstated conditions (\textit{W} = 47.5, \textit{p} = .48, \textit{d} = 0.27) and between the Constant-NotReinstated and Varied conditions (\textit{W} = 90.0, \textit{p} = .03, \textit{d} = 0.91) was observed. Similarly, no significant difference between the Constant-NotReinstated and Constant-Reinstated conditions (\textit{W} = 60.0, \textit{p} = .37, \textit{d} = 0.19) and between the Constant-NotReinstated and Varied conditions (\textit{W} = 58.0, \textit{p} = .25, \textit{d} = 0.67) was observed in the low-presence group.

The forgetting ratios for each presence group are presented in \cref{fig:forgetting_ratio-presence}. In the high-presence group, no significant difference between the Constant-NotReinstated and Constant-Reinstated conditions (\textit{W} = 53.0, \textit{p} = .28, \textit{d} = 0.73) and between the Constant-NotReinstated and Varied conditions (\textit{W} = 92.0, \textit{p} = .02, \textit{d} = 1.06) was observed. Similarly, no significant difference between the Constant-NotReinstated and Constant-Reinstated conditions (\textit{W} = 65.5, \textit{p} = .19, \textit{d} = 0.32) and between the Constant-NotReinstated and Varied conditions (\textit{W} = 55.0, \textit{p} = .38, \textit{d} = 0.56) was observed in the low-presence group.

\subsection{VR Sickness}
\begin{figure}[tb]
\centering
\includegraphics[width=0.7\columnwidth]{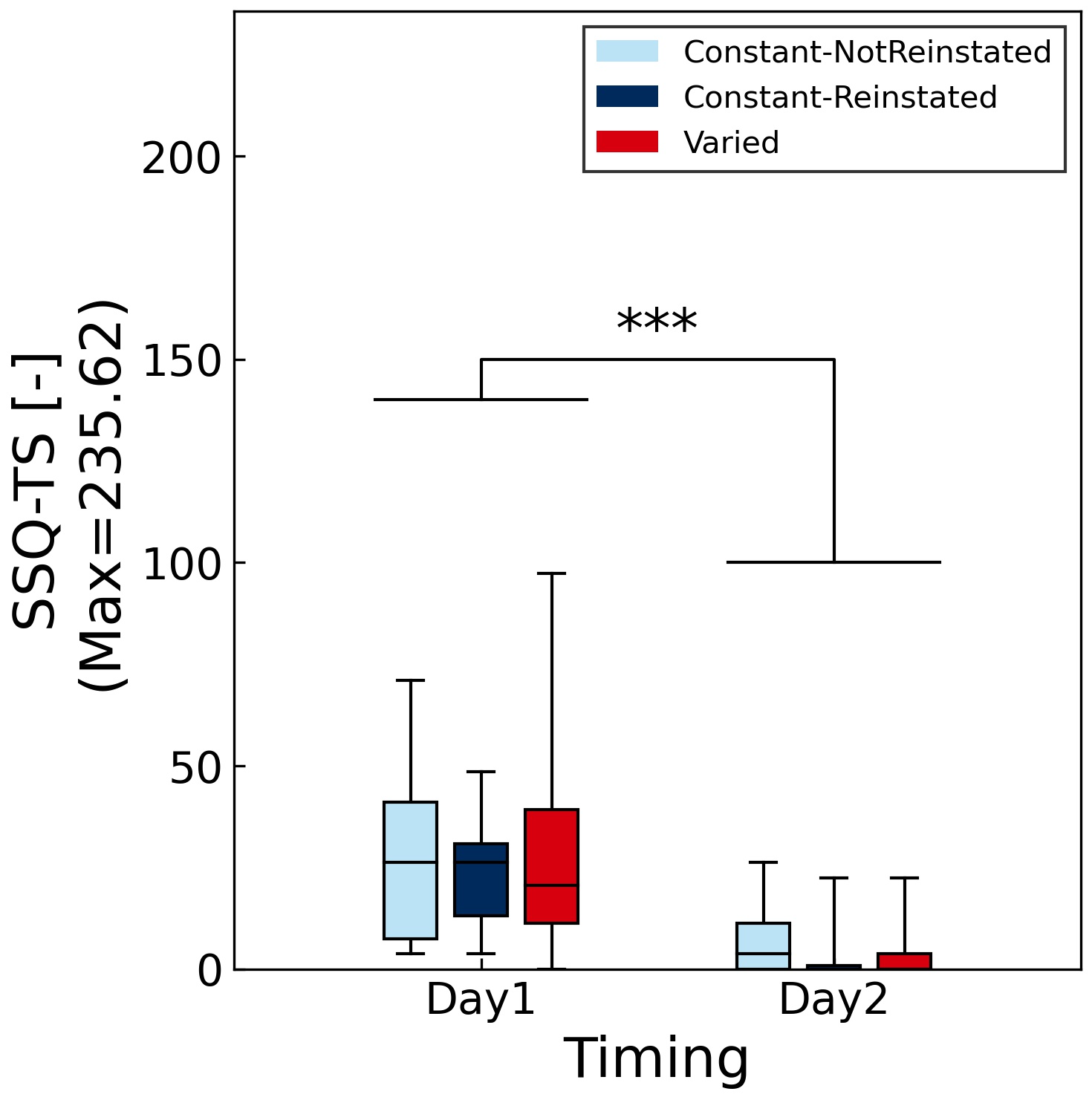}
\caption{SSQ total severity scores (SSQ-TS) for VR sickness. Sickness was significantly higher after Day 1 than after Day 2. No significant difference between the context conditions was observed. In addition, no correlation existed between VR sickness after Day 1 and memory performance measurements.}\label{fig:ssq}
\end{figure}

After conducting the experiment for two days, the participants answered SSQ to indicate the degree of VR sickness they felt.
List-wise, two participants were removed and excluded from the analysis because they were missing some values.
The SSQ total scores (SSQ-TS) obtained on Day 1 and Day 2 in each context condition are presented in \cref{fig:ssq}.
After aligned rank transformation~\cite{Wobbrock-etal2011}, three-by-two mixed ANOVA was computed, with context (Constant-NotReinstated, Constant-Reinstated, and Varied) as a between-participants variable and measurement timing (Day 1 and Day 2) as a repeated measure.
The result showed that only the main effect of measurement timing was significant (\textit{F} (1, 58) = 132.9, \textit{p} $<$ .001, ${\eta_p}^2$ = 0.70). Conversely, the main effect of context and the interaction were not significant (\textit{F} (2, 58) = 1.09, \textit{p} = .34, ${\eta_p}^2$ = 0.04; \textit{F} (2, 58) = 0.37, \textit{p} = .69, ${\eta_p}^2$ = 0.01).

We calculated Spearman's correlation between the SSQ-TS at Day 1 and RP1, retention, and forgetting to examine the possibility that VR sickness was a confounding factor for memory performance.
We found no significant correlations with RP1 ($\rho$ = -0.02, \textit{p} = .87), retention ($\rho$ = -0.18, \textit{p} = .17), nor forgetting ($\rho$ = -0.07, \textit{p} = .58; for ratio, $\rho$ = -0.09, \textit{p} = .48).
No correlation analysis was performed for the SSQ-TS on Day 2 because the scores were too low for consideration.

\section{Discussion}
\subsection{No reinstatement effect using IVE context}
In this study, we investigated the reinstatement effect on all memory measures (RP1, acquisition, retention, and forgetting) and found no significant effects.
Therefore, the hypotheses were not supported.
Whereas the previous studies used 3DCG IVEs~\cite{Rocabado-etal2022, Shin-etal2021, Walti-etal2019}, this study is the first to use live-action 360-degree video for IVE.
We assumed that increased realism and familiarity would improve the effectiveness of the environmental context, but the assumption was not supported.

For RP1 and acquisition, the hypothesis stated that the number of correct responses obtained would be higher in the same context conditions (Constant-NotReinstated and Constant-Reinstated conditions), where tests were conducted in the same context as the initial study than in the different context condition (Varied condition), where tests were conducted in a new context.
From \cref{fig:recall}, we observed no differences between the conditions, which were very different from those observed by Smith and Handy~\cite{Smith-Handy2014, Smith-Handy2016} who showed the superiority of the same context condition to the different context condition by video/photo context manipulation in a similar procedure.
Particularly in RP1, the number of correct answers \removed{tended to be slightly higher in the different context condition than in the same context condition}\added{was not significantly different between the same and different context conditions with the small effect size} (same context: \textit{M} = 3.71, \textit{SD} = 2.52; different context: \textit{M} = 4.05, \textit{SD} = 3.01; Cohen's \textit{d} = 0.12)\removed{, which was the opposite to the hypothesis. Thus, the result of no reinstatement effect observation was not obtained due to insufficient detection power}.

For retention, the hypothesis stated that the number of correct responses would be higher in the Constant-Reinstated condition, where participants performed the final test in the same context as Day 1, than in the Constant-NotReinstated condition, where the participants performed the final test in a context different from that of Day 1.
Similarly, for forgetting, the hypothesis stated that forgetting would be lower in the Constant-Reinstated condition than in the Constant-NotReinstated condition.
From \cref{fig:recall} and \cref{fig:forgetting}, we observed no significant differences between the conditions, and the hypothesis was not supported.
\removed{However, unlike in the case of RP1 and acquisition, the Constant-Reinstated condition was numerically dominant.}
\removed{Although the}\added{The small} effect size \removed{was small,}\added{suggests that} the detection power may have been insufficient.

The lack of effect may have been caused by cue overload~\cite{Rutherford2004}, which is a common phenomenon observed in this field of study. Although the use of too many words eliminates the reinstatement effect, no such occurrence was analyzed in our experiment because Smith and Handy~\cite{Smith-Handy2016}, who also used 20 Tagalog-English pairs of words, detected the effect.
On the other hand, some procedural differences were observed between our study and that of Smith and Handy~\cite{Smith-Handy2016}. For example, Smith and Handy quickly switched the background photo contexts (one photo per word) in their study, while we used the same context for 20 items. The experimental protocol was modified because quick changes in IVE could confuse the participants and compromise the sense of presence, which may be vital for inducing context-dependent effects using VR. This difference in cue load level was not problematic in that it avoided the problem of cue overload but may have reduced the reinstatement effect~\cite{Smith-Manzano2010}.
However, W\"{a}lti et al.~\cite{Walti-etal2019} already proved that even in a method like Smith and Handy's, where the contexts are changed quickly per word, no reinstatement effect could be obtained in context manipulation by VR.
Therefore, we do not consider that cue overload can explain the null results.

Instead, the unique features of VR may have weakened the effect.
From the discussion presented by Watson and Gaudl~\cite{Watson-Gaudl2021}, we deduced that the congruence of the meta-cognitive context of ``experiencing VR'' may have counteracted the effects of visual and auditory context changes.
This ``experiencing VR'' context was thought to be formed by wearing the HMD and high arousal during the VR experience.
If this assumption is correct, the context cannot be changed irrespective of the extent of manipulation of audio-visual environmental information in VR.
Similarly, the congruence of the meta-cognitive context of ``being at the same university and undergoing the same experiment'' was suggested to weaken the context-dependent effect when using the room as the environmental context~\cite{Bjork-RichardsonKlavehn1989}.
However, despite the meta-cognitive context match, ways in which the effect can be stably detected have been suggested, for example, by using the background video context~\cite{Smith-Manzano2010} or simultaneous manipulation of place, task, and social context~\cite{Isarida-Isarida2004}.
Moreover, as the decontextualization effect was confirmed in the present study, the meta-cognitive contextual congruence of experiencing VR is suggested not to be fatal.
Therefore, further research is expected to reveal conditions under which the reinstatement effect occurs using VR despite the meta-cognitive context match.

One method to solve the meta-cognitive consistency problem could employ a further increase in presence.
\removed{In this study, the group of participants who reported higher presence tended to show a more evident reinstatement effect than those who reported lower presence, as shown in \mbox{\cref{fig:interaction-presence-forgetting}}. Although this trend was not statistically significant, the fact that the prevailing conditions were switched is noteworthy. The high-presence group forgot less in the Constant-Reinstated condition, while the low-presence group forgot less in the Constant-NotReinstated condition.}
\added{For example, i}\removed{I}ncreasing body tracking levels may be considered effective for enhancing presence~\cite{Cummings-Bailenson2016}. As only three tracking points were used (the head and both hands) in this study, presence can be improved by attaching trackers to the waist and feet of the participants to achieve full-body tracking. Further, if the movements of the fingers can be faithfully reflected in the avatar by hand tracking, the presence can be improved.
\removed{However, the possibility of the relationship between presence and context-dependent effect indicated that participants, who more likely felt the presence, were more likely to be context-dependent because both characteristics are dependent on hippocampal activity~\mbox{\cite{Bird-Burgess2008, Cummings-Bailenson2016}}.
A test to manipulate presence as an independent variable should be conducted in a future study, considering its importance.}
\added{In this study, we did not observe a statistically significant impact of presence on reinstatement effects. However, if we continue our investigation with presence as the primary variable, we may observe an effect because both context-dependency of episodic memory and the sense of presence in IVE are dependent on hippocampal activity~\mbox{\cite{Bird-Burgess2008, Cummings-Bailenson2016}.}}

\subsection{Decontextualization effect using IVE context}
By comparing forgetting in the Varied and Constant-NotReinstated conditions, we proved a significant decontextualization effect, as shown in \cref{fig:forgetting}.
The effect size was also large. This study is the first to reveal the decontextualization effect using VR.

The two measures of forgetting showed significant effects, whereas no such observations were made for retention because the participants did not remember the maximum number of words during the five retrieval practices in the Constant-NotReinstated condition.
Smith and Handy~\cite{Smith-Handy2016} found that when using retention as a dependent measure, a significant effect was observed only when the participants performed five retrieval practices. No effect was observed when fewer retrieval practices were conducted. Thus, the participants have to remember almost all the words during the last retrieval practice to detect the decontextualization effect on retention.
In this study, the participant's performance on the fifth retrieval practice was approximately 70\%, which has to be increased.
Conversely, Smith and Handy~\cite{Smith-Handy2016} found that the effect was significant regardless of the number of retrieval practices when forgetting was the dependent variable.
These observations explain the present results well.

The background theory of the decontextualization effect includes encoding variability and desirable difficulty hypotheses.
Unlike the results of Smith and Handy~\cite{Smith-Handy2014, Smith-Handy2016}, the desirable difficulty hypothesis was not supported in this study because no significant difference between conditions in the acquisition was observed during the experiment.
The correctness of the two hypotheses can be clarified by examining the case where restudying would be used instead of retrieval practice.
If the encoding variability hypothesis is correct, the decontextualization effect should also be observed in the restudying protocol.

Watson and Gaudl~\cite{Watson-Gaudl2021} addressed the research question closest to our study. In their experiment, participants learned ten words in one VR room, moved through a virtual door in the IVE to another VR room, and learned another set of ten words. Thereafter, they hypothesized that recall performance improved compared to a condition where they did not pass through a door and the room did not change. However, this hypothesis could not be supported. The most notable difference was that they failed to find a significant context-dependent effect. Conversely, we identified a significant context-dependent effect in our study, but several other experimental methodological differences provided us with some critical perspectives.
The first deviation from this study is that they split a single wordlist and learned each sub-list once in different environmental contexts. Conversely, in this study, the environmental context was varied in which the same word list was repeatedly learned.
These two experimental methods have traditionally been discussed under the same category~\cite{Smith-Vela2001}, as they share the use of multiple environmental contexts. However, a question on their togetherness as the same phenomenon exists. The lack of examples from existing research contributes to the existence of these issues, which can be attributed to the high cost of using multiple environmental contexts in the past. Thus, VR has the potential to solve this unresolved problem by enabling the conduction of low-cost multiple-context experiments.

The second difference between our study and that of Watson and Gaudl~\cite{Watson-Gaudl2021} includes the type and number of IVEs used. Watson and Gaudl used two photorealistic 3DCG indoor environments, while we used seven 360-degree-video outdoor environments. Based on the encoding variability hypothesis, the more environmental contexts used, the more cues were available. Thus, the difference in the number of contexts might have improved the recall performance~\cite{Smith1982}. Additionally, the differences between 3DCG and 360-degree video, indoor and outdoor, can also be verified in the future.

Finally, Watson and Gaudl used doors for a contextual change, as doors have been found to act as boundaries for episodic memory~\cite{Radvansky-Copeland2006}. Conversely, the participants of this study remained seated and only put on and off their HMD. Therefore, the possibility that putting on and off the HMD acted as an episodic boundary similar to a door or visually representing movement (e.g., virtual doors or portals) in the IVE may have improved the effect remains to be explored.

\removed{The difference between the Constant-NotReinstated and Varied conditions tended to be larger for participants with higher presence, as shown in \mbox{\cref{fig:interaction-presence-forgetting}}. Although this trend was not statistically significant, we consider it to be an essential hint for further improving the memory support effect obtained in this study.}
\added{
Although we did not observe a statistically significant effect of presence on the decontextualization effect, we are expecting the presence to be a possible parameter for improving the memory support effect observed in this study.
}
According to the aforementioned discussion on the reinstatement effect, we can manipulate presence as an independent variable for validation in the future; for instance, by manipulating the tracking level.

Finally, the VR sickness measured by the SSQ did not differ significantly between conditions, as shown in \cref{fig:ssq}. Therefore, the characteristic was not confounded for memory performance. VR sickness is one of the critical factors to be considered during practical VR-based applications. Therefore, this study is noteworthy in that we measured VR sickness for the first time in environmental context-dependent memory research using VR and showed relatively low severity.

\subsection{Limitations}
The most important limitation of this study was that the controls were not fully rigorous because the experiment was conducted remotely.
Conducting remote experiments was unavoidable owing to the infection control measures adopted against COVID-19 at the time of the study.
The visual stimuli during the memory experiment were likely to have been controlled because the HMD completely covered the vision of the participants. Conversely, the auditory stimuli may have been insufficiently controlled because each participant used their own earphones or headphones, which may have been affected by external sounds.
However, the participants were informed to perform the experiment in a quiet place to keep such various situations under control.

Furthermore, a bias in the characteristics of the participants was observed. The number of men and women was matched across the conditions, but there were fewer female participants overall.
The effect of gender on the context-dependent effects is not yet clear. However, recruiting an equal number of representatives in future experiments would be desirable.
Additionally, limiting the participants to those who had their own HMD may have biased the study toward those familiar with VR.
As noted in Section~\ref{participants}, the degree of participants' previous VR experience averaged 2.87 on a 5-point scale ranging from 0 to 4.
This issue may not be critical in an era when VR is becoming more common, but it is worth considering in the current transitional period.

In considering the generalizability of this study, the retention interval length is an important parameter. We used a two-day retention interval between the learning and testing. This duration aligns with previous studies~\cite{Smith-Handy2014}, but further investigations using various lengths of retention interval would be practically important, such as for educational settings. Meta-analysis findings~\cite{Smith-Vela2001} indicate that context-dependent effects tend to exhibit larger effect sizes when longer intervals ranging from one day to one week are employed, compared to shorter intervals like five minutes. Furthermore, Smith and Handy~\cite{Smith-Handy2014} found significant decontextualization effects even with five-minute intervals. Considering these findings, we believe that our results have some generalizability regarding the length of the retention interval.

Additionally, this study demonstrated a new method of foreign language learning as a clear application by employing the task of memorizing Japanese-Tagalog pairs. However, the generalizability of this research findings may go beyond language learning. For example, Smith and Handy (2014) showed decontextualization effects in a face/name pair memory task~\cite{Smith-Handy2014}, and it is highly possible that our findings can be applied to various areas as long as classified in the field of episodic memory. On the other hand, the results are unlikely to be applicable to memories that cannot be described verbally, such as memories of motor skills. We employed language learning to be consistent with the use of word lists as the material to be memorized in most previous studies and also because Smith and Handy (2016)~\cite{Smith-Handy2016}, who found decontextualization effects, employed Tagalog-English pair learning. Future studies are expected to conduct validation in various situational settings to clarify the generalizability of the results.

\subsection{Future Work}
In this experiment, it was verified that while the reinstatement effect using VR is unlikely to occur, the decontextualization effect does occur.
As the decontextualization effect using VR was revealed for the first time in this study, replication and follow-up experiments should be conducted to elucidate the conditions under which the effect can be enhanced.
For example, \removed{the experiment in this study suggested that the context-dependent effect may become more apparent in participants with higher presence. T}\added{t}he relationship between presence and context-dependent effect is\removed{, therefore,} \added{an} essential topic for future work.

Additionally, verifying whether the effect can be improved by manipulating social context, such as who is in the audiovisual environmental context, is important.
Currently, the social context in VR can be manipulated more easily than earlier by placing virtual models in the CG world. For example, Sadeghi et al.~\cite{Sadeghi-etal2022} used VR to prepare five levels of crowding in subway train cars and examined the impact on participants' valence and time estimation. As manipulating the number of people around an environmental context is difficult, the use of VR offers the advantage of expanding the scope of environmental context-dependent memory research. Additionally, recent research that manipulated the appearance of others instead of the number of people suggests that the appearance of the instructor avatar can be manipulated as an environmental context in remote classes to provide memory support effects~\cite{Mizuho-etal2023}. As the use of VR SNS is also expanding, the effects of the social context within VR should be investigated.

Finally, although the final test was conducted in an IVE, future research should investigate whether the same effect can be observed when the test is conducted in a physical environment.
Context-dependent forgetting has been shown to occur when the environmental context changes from IVE to the physical environment~\cite{Lamers-Lanen2021}. The inability to use what is learned within VR when returning to reality hinders the utilization of VR, which is a critical problem to be solved for various VR applications, such as education. 
Mizuho et al.~\cite{Mizuho-etal2023a} tested the hypothesis, which states that by increasing the visual fidelity of IVE, context-dependent forgetting reduces. However, no valuable results have yet been obtained till date.
Based on the results of this study, variations of IVEs during encoding for decontextualization would be a promising solution and a clue to promote the use of VR in the future.

\section{Conclusion}

This study tested whether context-dependent forgetting can be eliminated by inducing reinstatement and decontextualization effects using 360-degree video-based virtual environments. For the reinstatement effect, complex results were obtained, suggesting the need for further elaborate research. On the other hand, for the decontextualization effect, this study was the first to validate the effect using VR and revealed its statistically significant effect with a large effect size. While previous studies have assumed that VR could naturally reproduce the context-dependent effects observed in the real world, most of them could not prove the assumption. Given this context, obtaining a clear decontextualization effect in this study sheds light on the possibility of reproducing context-dependent effects using VR.

\section*{Acknowledgments}
This work was partially supported by JST Moonshot Research \& Development (JPMJMS2013); Grant-in-Aid for Challenging Research (Exploratory) (21K19784); Grant-in-Aid for JSPS Fellows (23KJ0422);
and ``Multimodal XR-AI platform development for tele-habitation and reciprocal care coupling with health guidance'' project (JPNP21004) subsidized by NEDO.

\bibliographystyle{IEEEtran}
\bibliography{reference}

\begin{IEEEbiography}[{\includegraphics[width=1in,height=1.25in,clip,keepaspectratio]{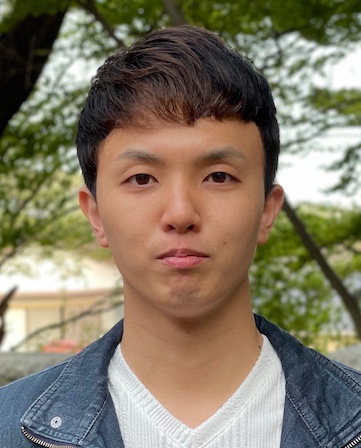}}]{Takato Mizuho}
is a Ph.D. student at the Graduate School of Information Science and Technology, the University of Tokyo. He received his B.S. and M.S. degrees from the University of Tokyo in 2020 and 2022, respectively. His research interests include avatars, context-dependent memory, presence, and virtual reality.
\end{IEEEbiography}

\begin{IEEEbiography}[{\includegraphics[width=1in,height=1.25in,clip,keepaspectratio]{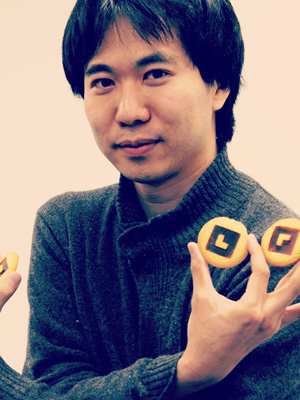}}]{Takuji Narumi}
is an associate professor at the Graduate School of Information Science and Technology, the University of Tokyo. He received his BE and ME degrees from the University of Tokyo in 2006 and 2008, respectively. He also received his Ph.D. in Engineering from the University of Tokyo in 2011. His research interests broadly include perceptual modification and human augmentation with virtual reality and augmented reality technologies.
\end{IEEEbiography}

\begin{IEEEbiography}[{\includegraphics[width=1in,height=1.25in,clip,keepaspectratio]{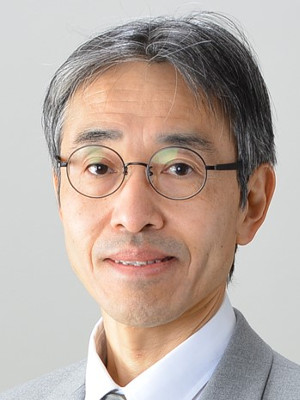}}]{Hideaki Kuzuoka}
is a professor at the Graduate School of Information Science and Technology, the University of Tokyo. He graduated from the Graduate School of Engineering, the University of Tokyo, in 1986 and received his Ph.D. in Engineering in 1992. His research interests include computer-supported cooperative work, social robotics, virtual reality, and human-computer interaction in general.
\end{IEEEbiography}

\end{document}